\documentstyle[aps,prl,twoside,multicol,epsf,psfrag,graphicx]{revtex}
\begin{document}
\draft

\title{Enhancement of mobility by periodically
modulating the slanting slope of a washboard potential}
\author{Mangal C. Mahato$^\dagger$ and A.M. Jayannavar\cite{AMJ}}
\address{$^\dagger$Department of Physics, Guru Ghasidas University,
Bilaspur 495 009, Chattisgarh, India.}
\address{$^\star$Institute of Physics, Sachivalaya Marg,
Bhubaneswar-751005, India}

\maketitle

\begin{abstract}

Average mobility of very feebly damped particles in tilted periodic
potentials is considered. Under the combined action of thermal
fluctuations and small temporal modulation of the tilt of the potential
the particles , in the small tilt range, become more mobile than 
without modulation. The enhancement of mobility depends (nonmonotonically)
on the frequency of modulation. For small modulations the enhancement 
shows a peak as a function of frequency. This has an obvious implication 
on the measured voltage across a Josephson junction driven by a small amplitude
alternating current of suitable frequency.
\end{abstract}

\thispagestyle{empty}

\pagestyle{myheadings}
\markboth{{\it Mangal C. Mahato and A.M. Jayannavar}}{
{\it Enhancement of mobility by periodically modulating ...}}

\begin{multicols}{2}

The inertia of a particle plays spectacular role when the motion is very
feebly damped even when the effective applied force is very small. The
temperature of the system helps overcome the effect of damping on the 
motion significantly. The present work concerns motion of such particles
in a periodic potential system of period $2\pi/k$ and amplitude $V_0$. When 
a tilt as small as $F_0 = 0.07V_0k$ is given to the potential at a 
temperature $T = 0.4V_0/k_B$, where $k_B$ is the Boltzmann constant,
 the particles make several, at times hundreds, of crossings across the
potential peaks before making a "halt" in a potential well. The particles 
start and continue
the journey (on the average downhill) again assisted by thermal fluctuations.
The intervals of these continued motions and halts (that is, the intervals
of the particles being in running and locked states, respectively) are not 
regular but must
follow broad distributions. A more or less similar picture appears when the
potential tilt is modulated periodically in time with amplitude $\Delta F=
0.01V_0k, 0.02V_0k, .. etc$ (Fig. 1). It has been shown earlier that upon
modulation the mobility of the particle exhibits hysteretic behavior. The
condition of criticality of such hysteresis loops and corresponding jump
distributions, which show transition between exponential to power law
behavior, have been discussed in reference\cite{Mar}. It is also stated 
there that
no appreciable hysteresis loop could be obtained if the average tilt $F_0$ is
chosen relatively far from a threshold value dependent on the product of
the friction coefficient $\gamma_0$ and $V_0$. We find that if the friction
coefficient becomes space dependent, and in particular periodic with the same
periodicity $2\pi/k$ but with a phase shift $\phi$ and amplitude $\lambda$,
appreciable hysteresis loops can be obtained for values of $F_0$ roughly in 
the range
between $0.05V_0k$ and $0.14V_0k$. 
The periodically varying friction
coefficient corresponds analogously to the term representing interference
between the quasiparticle tunneling and the Cooper pair tunneling across
Josephson junctions\cite{Falco}. The
inclusion of nonuniform friction coefficient does not give qualitatively
different result. However, quantitatively it makes difference. We, therefore, 
consider space dependent friction coefficient in our study.

\begin{figure}
\psfrag{x}{{\large $x$}}
\psfrag{t}{{\large $t$}}
\psfrag{lambda0df0}{{\tiny $\lambda=0,\Delta F=0$}}
\psfrag{lambda.9df0}{{\tiny $\lambda=.9,\Delta F=0$}}
\psfrag{lambda0df.01}{{\tiny $\lambda=0,\Delta F=.01$}}
\psfrag{lambda.9df.01}{{\tiny $\lambda=.9,\Delta F=.01$}}
\protect\centerline{\epsfxsize=3.2in \epsfbox{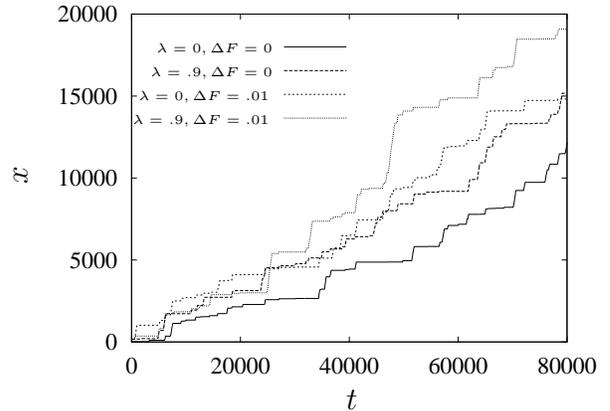}}
\caption{Distance $x(t)$ travelled in time $(t)$. The periodicity
$T_\Omega$ of modulation is 8000. Note that the periodicity of the
potential is only $2\pi$. (All quantities are in dimensionless
units; for example $kx$ is plotted as $x$.) Even the small vertical
steps correspond to the particle travelling for hundreds of periods
of the potential.}

\end{figure}

Consider the equation of motion\cite{Mah1},
\begin{equation}
m\frac{d^2x}{dt^2}+\gamma(x)\frac{dx}{dt}-V_0k cos(kx)
-F(t)-\surd(k_BT\gamma(x)){\hat f(t)}=0,
\end{equation}
where $\gamma(x)=\gamma_0(1-\lambda sin(kx+\phi))$ is the space dependent 
friction coefficient and $V(x)=-V_0 sin(kx)$ is the potential profile. The
external force has two parts one giving the average tilt to the potential and
other giving the time periodic modulation: $F(t)=F_0+\Delta F cos(\Omega t)$.
The fluctuating term $\hat f(t)$ satisfies $\left< \hat f(t) \right>=0$ and
$\left< \hat f(t)\hat f(t') \right>=2\delta (t-t')$ and may represent thermal
noise at temperature $T$. Equation (1) is exactly the same as the Josephson 
junction equation except that one needs to replace $\phi$ 
by $\pi/2$\cite{Falco,Ambe}.
In our calculations we take $\lambda=0.9$ (incidentally, the
choice happens to be in conformity with the experimental value\cite{Falco1})
and for comparison $\lambda=0$ (corresponding to uniform friction).

We solve the Langevin equation (1) (in fact, in its dimensionless form) 
numerically\cite{Nume} to calculate the average mobility $\mu$ 
($=\lim_{t\rightarrow\infty}(\frac{x(t)}{t})/F_0$) as a function of $F_0$ for various values of 
$\Delta F$ and $\Omega = \frac{2\pi}{T_\Omega}$, $T_\Omega$ is the periodicity
of the external field $F(t)$. Here $x(t)$ is the distance travelled by the 
particle in time $t$\cite{Foot1}.

Fig. 2 shows the plot of $\gamma_0\mu$ versus $F_0$ for $\Delta F=0$, 
$0.01V_0k$, $0.02V_0k$ and $0.04V_0k$ and choosing $T_\Omega$ in all these 
cases such that $(\frac{\Delta F}{V_0k})/(\frac{T_\Omega}
{m^{.5}V_0^{-.5}k^{-1}})$=constant=0.00001 (i.e., with a constant rate of 
change of the external field). The interesting aspect to note in the figure,
which is also the main result of this paper, is that for a range of values 
of small $F_0$ (the range depending on the value of $\Delta F$) the average 
mobility
is larger for $\Delta F\ne 0$ than for $\Delta F=0$. The result indicates that 
the mobility has been enhanced by modulating the slanting (small $F_0$, smaller
than the known critical values\cite{Mar}) slope
of the periodic potential. In other words, the effective friction could be
reduced below the mean friction just by modulating the potential periodically.
It is worth mentioning that in tribology such problems are often sought to
address and sometimes similar solutions hinted at\cite{Ptoday}. It is also
to be noted from the figure that the enhancement of mobility is larger for
larger $\frac{\Delta F}{F_0}$. However, we restrict our attention to 
$\frac{\Delta F}{V_0k}$ = 0.01 and 0.02 which are relatively quite small
compared to $\frac{F_0}{V_0k}$ of our interest where we expect mobility
enhancement.

\begin{figure}
\psfrag{gammamu}{{\large $\gamma_0\mu$}}
\psfrag{f0}{{\large $F_0$}}
\psfrag{deltaf0}{{\tiny $\Delta F=0$}}
\psfrag{deltaf.01}{{\tiny $\Delta F=.01$}}
\psfrag{deltaf.02}{{\tiny $\Delta F=.02$}}
\psfrag{deltaf.04}{{\tiny $\Delta F=.04$}}
\protect\centerline{\epsfxsize=3.2in \epsfbox{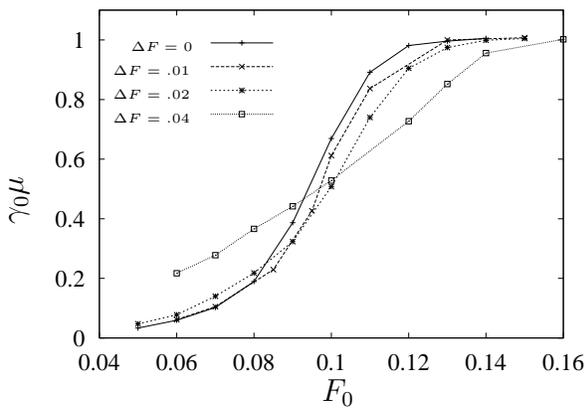}}
\caption{$\gamma_0\mu$ versus $F_0$ for various values of $\Delta F$.
The periodicity $T_\Omega$ for $\Delta F=.01$ is taken to be 1000,
for $\Delta F=.02$ it is 2000, and for $\Delta F=.04$ it is 4000 so
that the rate of field sweep is the same for all the three cases.
(All quantities in dimensionless units)}
\end{figure}

The occurrence of the mobility enhancement can mainly be attributed to the 
nonlinear variation of $\gamma_0\mu$ as a function of $F_0$ with 
$\Delta F = 0$. As can be seen from the figure the enhancement occurs where 
the curvature of the $\gamma_0\mu$-$F_0$ plot is concave upward. And this 
happens at lower values of $F_0$. At low values of $F_0$, during the positive
half cycle of the modulating field $F(t)$ the system samples more higher
velocities than it samples lower values during the negative half cycle
resulting in a positive gain in the average mobility. The situation is just the
reverse for higher values of $F_0$. The explanation is supported by the shapes
of the hysteresis loops (Fig. 3) obtained because of modulation. As can be seen
from the figure, for example, the hysteresis loop for $F_0=0.8V_0k$ is more
oval upward right whereas for $F_0=0.11V_0k$ the loop is more oval downward 
left. However, the shapes of the loops themselves are determined by the
distribution of intervals of locked and running states of the particles. And
the distributions are sensitive to the values of $F_0$ and other parameters 
such as the amplitude and frequency of modulation, etc. The choice of 
frequency of 
modulation is very important for mobility enhancement.

\begin{figure}
\psfrag{gammamu}{{\large $\gamma_0\mu$}}
\psfrag{force}{{\large $F$}}
\psfrag{deltaf0}{{\tiny $\Delta F=0$}}
\psfrag{f0eq.06}{{\tiny $F_0=.06$}}
\psfrag{f0eq.08}{{\tiny $F_0=.08$}}
\psfrag{f0eq.10}{{\tiny $F_0=.10$}}
\psfrag{f0eq.11}{{\tiny $F_0=.11$}}
\psfrag{f0eq.12}{{\tiny $F_0=.14$}}
\psfrag{f0eq.14}{{\tiny $F_0=.14$}}
\protect\centerline{\epsfxsize=3.2in \epsfbox{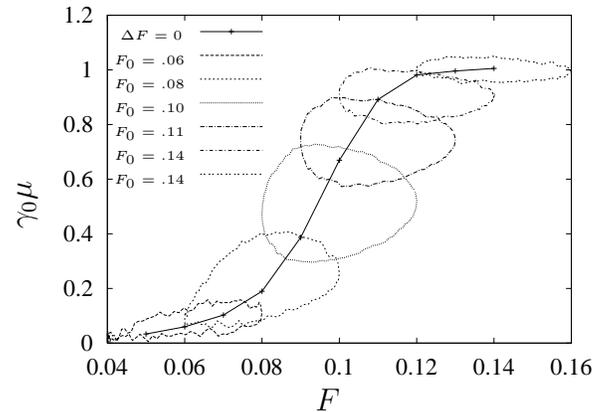}}
\caption{$\gamma_0\mu$ is plotted as a function of $F_0$ (solid curve).
The hysteresis curves are for various constant values of $F_0$ and
$\Delta F=.02$ with period of modulation $T_\Omega=2000$. (All
quantities in dimensionless units.)}
\end{figure}

Figure 4 shows the variation of average mobility as a function of frequency
$\Omega (=1/T_\Omega)$ of modulation. For large frequencies the mobility 
is small
and close to the value for the one without modulation (shown by the horizontal
line in the figure). As the frequency is decreased the mobility increases, 
attains a maximum and thereafter tends to decrease. The maximum enhancement
 for $\Delta F=0.01V_0k$ is about 10 per cent
whereas for $\Delta F=0.02V_0k$ it is as large as about 70 per cent of the
mobility obtained without modulation\cite{Git}. 
For $\Delta F=0.01V_0k$ the
curve has a distinct peak. However, for $\Delta F=0.02V_0k$ the peak is not as 
clear. For small modulation, as the frequency goes toward zero $(\Delta F\ne 0,
\Omega\rightarrow 0)$ the mobility decreases and tends to become closer to
the one corresponding to without modulation $(\Delta F=0)$. But for relatively 
large modulation (e.g., $\Delta F=0.02V_0k$, for $F_0=0.07V_0k$) the same does 
not seem to be true\cite{Foot2}. Therefore, for relatively small 
amplitude modulation it
is possible to maximize the enhancement of mobility by properly tuning the 
frequency of modulation. In the Josephson junction parlance it is reasonable
to state that the sensitivity of ac SQUIDs can be improved by properly
choosing the amplitude and frequency of applied alternating 
current\cite{Foot3}.

\begin{figure}[b]
\psfrag{gammamu}{{\large $\gamma_0\mu$}}
\psfrag{Omega}{{\large $\Omega$}}
\protect\centerline{\epsfysize=3.2in \epsfxsize=3.2in \epsfbox{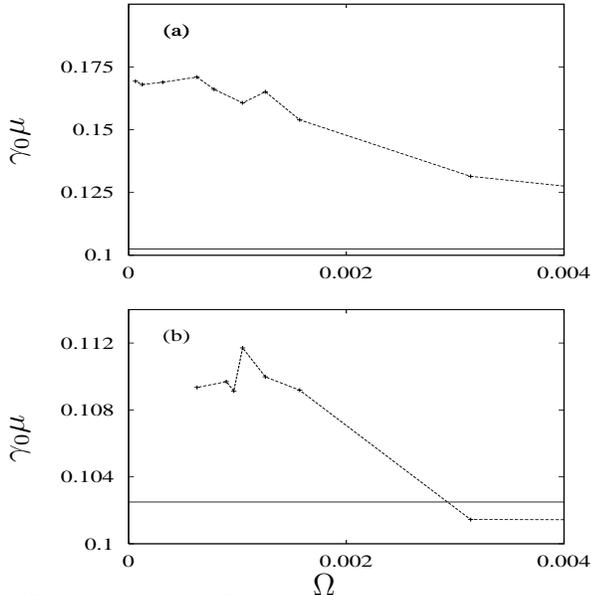}}

\caption{$\gamma_0\mu$ versus $\Omega$, the frequency of modulation
with $F_0=.07$ and (a) $\Delta F=.02$ and (b) $\Delta F=.01$. The
horizontal line is drawn for reference corresponding to $\gamma_0\mu$
for $\Delta F=0$. (All quantities in dimensionless units.)}
\end{figure}

In the entire process of particle motion temperature plays a very important 
role. At zero temperature, for example, once the particle gets trapped at any
one of the troughs of the potential, given the values of $F_0$ and $V_0$ 
considered, it will remain locked for ever making the particle effectively
immobile. It is the thermal fluctuations that release the particle from the 
locked states into the running states\cite{Risken}. As one can see from Fig. 1b
of reference\cite{Mar} that the mobility versus $F_0$ curve (for small $F_0$)
changes very rapidly with temperature. Their curvatures at small $F_0$ 
too differ. One would thus expect that mobility enhancement for given values 
of $F_0$,
 $\Delta F$ and $T_\Omega$ will show a peaking behavior showing the 
phenomenon of stochastic resonance as a result of synchronization 
due to the interplay between external drive and thermally assisted running to
locked transitions\cite{SR}. This aspect is presently being 
investigated.

\vspace{1.0cm}
\begin{center}
{\Large{\bf ACKNOWLEDGEMENT}}
\end{center}
MCM thanks the Institute of Physics, Bhubaneswar for partial support and
hospitality. The authors thank BRNS, DAE, India for financial support.

\end{multicols}

\begin{thebibliography}{999}
\bibitem{Foot1}
To be exact we should have taken a large number of realizations of $x(t)$
to get the average mobility. However, since $t$ is very large one realization
of the trajectory is assumed to be quite reasonable and adequate for
calculating the average.
\bibitem{Foot2}
For a reasonably good average of mobility one needs to calculate $x(t)$
for about 5000 cycles of modulations. For scaled $T_\Omega=6000$ the 
corresponding scaled times 
are typically of the order of $10^7$ to $10^8$. The CPU time needed
by a PII machine is more than 15 days. For larger values of $T_\Omega$ the
averaging costs enormous computer time and is neither practical nor feasible,
 with facilities at our disposal, to get results for smaller frequencies.
And for $\Delta F=0.02V_0k$ to see the peak, if at all it has one, one
either have to employ a better numerical method for the solution or a faster
machine.
\end{thebibliography}
\end{document}